\documentclass[twocolumn,showpacs,preprintnumbers]{revtex4}
\usepackage{bm}
\usepackage{amssymb}
\usepackage{amsmath}
\usepackage{graphicx}
\begin{document}
\title{Phase separation in the crust of accreting neutron stars}
\author{C. J. Horowitz}\email{horowit@indiana.edu}
\affiliation{Department of Physics and Nuclear Theory Center,
             Indiana University, Bloomington, IN 47405}
\author{D. K. Berry}\email{dkberry@indiana.edu}
\affiliation{University Information Technology Services,
             Indiana University, Bloomington, IN 47408}
\author{E. F. Brown}\email{ebrown@pa.msu.edu}
\affiliation{Department of Physics and Astronomy, National Superconducting Cyclotron Laboratory and Joint Institute for Nuclear Astrophysics, Michigan State University, East Lansing, MI 48824}             

\date{\today}
\begin{abstract}
Nucleosynthesis, on the surface of accreting neutron stars, produces a range of chemical elements.  We perform molecular dynamics simulations of crystallization to see how this complex composition forms new neutron star crust.  We find chemical separation, with the liquid ocean phase greatly enriched in low atomic number elements compared to the solid crust.  This phase separation should change many crust properties such as the thermal conductivity and shear modulus.  
\end{abstract}
\smallskip
\maketitle

\section{Introduction}
Phase separation is important for white dwarf stars \cite{wdwarf}.   As a star cools, and crystallization takes place, the crystal phase is enriched in oxygen while the liquid is enriched in carbon.  We believe phase separation may also be important for neutron stars that accrete material from a companion.  This material can undergo nuclear reactions involving rapid proton capture (the rp process) to synthesize a variety of medium mass nuclei \cite{rpash}.  Further accretion increases the density of a fluid element until crystallization occurs.  However, as we explicitly demonstrate with molecular dynamics simulations, crystallization is accompanied by chemical separation.  The composition of the new solid crust is very different from the remaining liquid ocean.  This changes many properties of the crust and can impact many observables.  

\begin{figure}[ht]
\begin{center}
\includegraphics[width=2.75in,angle=0,clip=true] {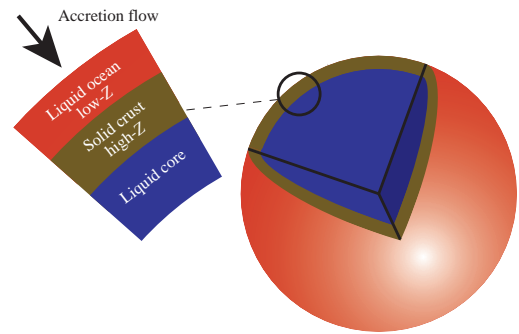}
\caption{(Color on line) Schematic diagram of the surface of an accreting neutron star.  This paper focuses on chemical separation upon crystallization at the boundary between the liquid ocean and the solid crust.  We find that the ocean is enriched in low $Z$ elements.  Note that the boundary between the inner and outer crust is not shown.} 
\label{Fig0}
\end{center}
\end{figure}

With chemical separation, the liquid ocean, see Fig. \ref{Fig0}, is greatly enriched in low atomic number $Z$ elements.    Carbon, if present, may be depleted in the crystal (crust) and enriched in the liquid ocean phase.   Also, chemical separation may change the thermal conductivity of the crust and its temperature profile.  Indeed some neutron stars are observed to produce energetic X-ray bursts known as superbursts.  These are thought to involve the unstable thermonuclear burning of carbon \cite{superbursts, superbursts2, superbursts3}.  However it is unclear how the initial carbon concentration is obtained and how the ignition temperature is reached. 

Chemical separation may significantly change the thickness, shear modulus, and breaking strain of the crust.  This could change the shape of a neutron star and the radiation of periodic gravitational waves \cite{jones2005} \cite{haskell}.  Furthermore, changes in the crust could change the properties of quasi periodic oscillations that may be observable in thermonuclear bursts.  

The process we consider here is distinct from sedimentation, which may also occur at lower densities where the ions are fluid \cite{peng}.   However, we are not aware of any previous calculations of chemical separation from crystallization for accreting neutron stars.  Jones has considered how a range of compositions may change the properties of the crust of a non-accreting neutron star.  In ref. \cite{jones88} he considers phase separation, but only based on quite early work on the free energy of the two-component  Coulomb plasma.  Instead, Jones suggests that the system will form an amorphous solid \cite{jones}.  

The ash resulting from rapid proton capture (rp process) nuclear reactions is expected to have a complex composition involving a number of different chemical elements \cite{rpash, rpash2}.  Unfortunately, it can be difficult to construct the phase diagram for a multi-component system.  The pure one component plasma (OCP) phase diagram is well known.  The liquid solidifies when the ratio of a typical Coulomb energy to the thermal energy $kT$ is $\Gamma \approx 175$ \cite{ocp}.  The parameter $\Gamma$ is defined,
\begin{equation}
\Gamma=\frac{Z^2 e^2}{a T},
\label{gamma}
\end{equation}
where the ion charge is $Ze$, the temperature is $T$, and the ion sphere radius $a$ describes a typical distance between ions, $a=(3/4\pi n)^{1/3}$.  Here $n$ is the ion (number) density.  The phase diagram for binary mixtures has also been determined.  See for example \cite{cophasediagram}.  For binary mixtures, the solid phase is enriched in the high $Z$ ion and the liquid phase is enriched in the low $Z$ ion.  

Often, the theoretical phase diagram is constructed from extremely accurate calculations of the free energies of the solid and liquid phases.  The melting point is determined by equating these two free energies.  Very accurate calculations are needed because the free energies are nearly parallel as a function of $T$.  A small error in the free energy of one phase can lead to a large error in the melting point.  Therefore, it may be very difficult to compute the free energy of multi-component systems with enough accuracy to determine the phase diagram.

Instead, in this paper we determine the phase diagram of a multi-component system directly via molecular dynamics (MD) simulations.  Our simulation volume contains regions of both the liquid and solid phase.  This approach has many advantages.  It is simple and robust.  Delicate free energy calculations are not needed.  One can directly measure the composition of the two phases that are in equilibrium.  Furthermore, one can run simulations with arbitrarily complicated compositions.

However, there are two limitations to direct molecular dynamics simulation.  First, finite size effects may be significant because a large fraction of the ions are near the interfaces between the two phases.  We minimize finite size effects by using a moderately large number of ions 27,648 and we measure the composition of the two phases in regions that are away from the interfaces.  Second, it can take a long time for the two phase system to come into thermodynamic equilibrium.  We address non equilibrium effects by running for a total simulation time of 151 million fm/c (over six million MD time steps) and by monitoring the time dependence of the composition of the two phases.  Still, as we discuss below, the system may not be in full equilibrium and this may be an important question for further work.  Nevertheless, we start with equal compositions and find dramatically different compositions for the liquid and solid, where the difference has increased with simulation time.  If the system is not in equilibrium by the end of our simulation, we expect the difference between liquid and solid to only increase further with time.  Therefore, we do not think non-equilibrium effects will change our conclusion that {\it the liquid and solid have very different compositions}. 

This paper is organized as follows.  In Section \ref{mdsimulation} we describe our molecular dynamics simulation.  Results are presented in Section \ref{results} and we conclude in Section \ref{conclusions}.

\section{Molecular Dynamics Simulation}
\label{mdsimulation}

We now describe the initial composition for our simulation.  Schatz et al. have calculated the rapid proton capture (rp) process of hydrogen burning on the surface of an accreting neutron star \cite{rpash}.  This produces a variety of nuclei up to atomic masses $A\approx 100$.  Gupta et al. \cite{gupta} then calculate how the composition of the rp process ash evolves because of electron capture and light particle reactions as the material is buried by further accretion.  Their final composition, at a density of $2.16\times 10^{11}$ g/cm$^3$ (near neutron drip at the bottom of the outer crust) has forty percent of the ions with atomic number $Z=34$, while an additional 10 \% have $Z=33$.  The remaining half of the ions have a range of lower $Z$ from $Z=8$ to 32.  Finally there is a small abundance of $Z=36$ and $Z=47$.
      
\begin{figure}[ht]
\begin{center}
\includegraphics[width=2.75in,angle=270,clip=true] {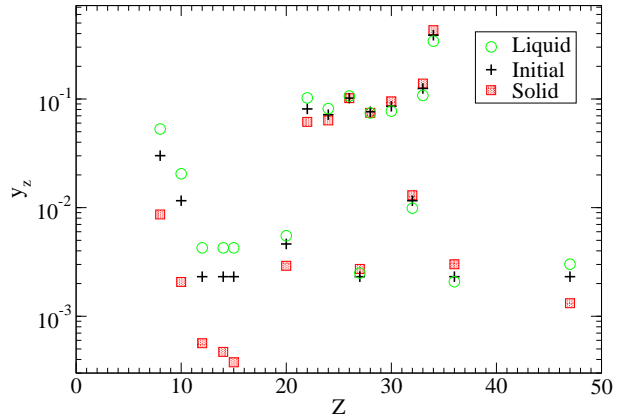}
\caption{(Color on line) Abundance (by number) of chemical elements versus atomic number $Z$.  The plus symbols show the initial composition of the mixture.  The final compositions of the liquid phase, open green circles, and solid phase, filled red squares, are show after a simulation time of $151\times 10^6$ fm/c, see Section \ref{results}.} 
\label{Fig1}
\end{center}
\end{figure}

For simplicity we use the Gupta et al. abundances because we have them available.  However these abundances were calculated assuming no phase separation.  Therefore they have not been determined self-consistently if there is phase separation.  Nevertheless, we use them to provide a first orientation.  
Note that we use abundances calculated near $10^{11}$ g/cm$^3$, while the ocean/ crust boundary may be near $10^{10}$ g/cm$^3$.  The differences in composition at these two densities may be primarily do to a modest amount of electron capture.  This should not significantly change our results.  Perhaps phase separation will lead to more important changes in the abundances.  Chemical separation is expected to change compositions over a large range of densities in addition to densities near the ocean/ crust interface.  For example, changes in composition of the liquid, near the crust interface, are expected to diffuse throughout the ocean.  As we discuss in section \ref{conclusions} future calculations of abundances including phase separation would be very useful.  
 
\begin{table}
\caption{Abundance $y_z$ (by number) of chemical element $Z$.  Results are presented for the original mixture and for the final liquid and solid phases after a simulation time of $151\times 10^6$ fm/c, see text.} 
\begin{tabular}{llll}
$Z$ & Mixture & Liquid & Solid \\
8 & 0.0301 & 0.0529 & 0.0087 \\
10 & 0.0116 & 0.0205 & 0.0021 \\
12 & 0.0023 & 0.0043 & 0.0006 \\
14 & 0.0023 & 0.0043 & 0.0005 \\
15 & 0.0023 & 0.0043 & 0.0004 \\
20 & 0.0046 & 0.0055 & 0.0029 \\
22 & 0.0810 & 0.1024 & 0.0616 \\
24 & 0.0718 & 0.0816 & 0.0635 \\
26 & 0.1019 & 0.1065 & 0.1017 \\
27 & 0.0023 & 0.0025 & 0.0027 \\
28 & 0.0764 & 0.0744 & 0.0746 \\
30 & 0.0856 & 0.0773 & 0.0949 \\
32 & 0.0116 & 0.0099 & 0.0130 \\
33 & 0.1250 & 0.1079 & 0.1388 \\
34 & 0.3866 & 0.3408 & 0.4297 \\
36 & 0.0023 & 0.0021 & 0.0030 \\
47 & 0.0023 & 0.0030 & 0.0013 \\
\end{tabular} 
\label{tableone}
\end{table}

As an initial composition we chose 432 ions with $Z$ and mass number $A$ drawn at random according to the Gupta et al. abundances.  This is shown in Fig. \ref{Fig1} and listed in Table \ref{tableone} and closely approximates the original distribution up to limitations of the small statistics.  We chose such a small system, 432 ions, to simplify producing the original solid configuration, see below.  Note that the liquid and solid phase results shown in Fig. \ref{Fig1} will be discussed in Section \ref{results}.

At these densities, electrons form a relativistic degenerate Fermi gas.  The ions are fully pressure ionized and interact with each other via screened Coulomb interactions.  The potential between the $i$th and $j$th ion is assumed to be,
\begin{equation}
v_{ij}(r)=\frac{Z_iZ_j e^2}{r} {\rm e}^{-r/\lambda}.
\label{v(r)}
\end{equation}
Here the ion charges are $Z_i$ and $Z_j$, $r$ is their separation and the electron screening length is $\lambda$.  For cold relativistic electrons, the Thomas Fermi screening length is $\lambda^{-1}=2\alpha^{1/2}k_F/\pi^{1/2}$ where the electron Fermi momentum $k_F$ is $k_F=(3\pi^2n_e)^{1/3}$ and $\alpha$ is the fine structure constant.  Finally the electron density $n_e$ is equal to the ion charge density, $n_e=\langle Z\rangle n$, where $n$ is the ion density and $\langle Z\rangle$ is the average charge.  Note that we are interested in temperatures near the melting point where the ion thermal de Broglie wave length is much shorter than the inter ion spacing.  Therefore quantum corrections to the ion motion should be very small.

We now describe the initial conditions for our classical MD simulation.  It can be difficult to obtain an equilibrium crystal configuration for a large system involving a mixture of ions.  Therefore, we start with a very small system of 432 ions with random coordinates at a high temperature and cool the system a number of times by re-scaling the velocities until the system solidifies.  Here the velocities of all of the ions are multiplied by a common factor so that the kinetic energy per ion is $3T/2$ for a series of decreasing temperatures $T$.  Next four copies of this solid configuration were placed in the top half of a larger simulation volume along with four copies of a 432 ion liquid configuration.  The resulting system with 3456 ions was evolved in time until it fully crystalized.  Finally, four copies of this 3456 ion crystal were placed in the top half of the final simulation volume along with four copies of a 3456 ion liquid configuration.  This final system has 27648 ions and consists of a solid phase above a liquid phase.  Note that the initial compositions of these two phases are equal.

Our results can be scaled to different densities.  For historical reasons, our simulation was run at a relatively high ion density of $n=7.18\times 10^{-5}$ fm$^{-3}$.  This corresponds to a mass density of $1.04\times 10^{13}$ g/cm$^3$.  However, this density can be scaled to any desired value $\hat n$ by also changing the temperature $\hat T$ so that $\hat n/\hat T^3=7.18\times 10^{-5}/(0.34360)^3$ (MeV-fm)$^{-3}$ this insures the value of $\Gamma$, see Eqs. \ref{gamma},\ref{gammamix}, remains the same.    Note that our simulations depend on the electron screening length $\lambda$ only in the ratio of $\lambda/a$.  For relativistic electrons, this ratio is independent of density.  Therefore, the above scaling works even with electron screening effects.  This is because the only length scale in the problem for both electron and ion interactions is related to $n^{-1/3}$.  

Many run parameters are collected in Table \ref{tabletwo}.  We evolve the system in time using the simple velocity Verlet algorithm \cite{verlet} with a time step $\Delta t=25$ fm/c.  We use periodic boundary conditions.  Our simulation volume is large enough so that the box length $L=727.5$ fm is much larger than the electron screening length $\lambda$.  Indeed $L/2\lambda=13.9$.  The screened potential, for two ions separated by a distance $L/2$, is very small.  This helps to reduce finite size effects.  We include the interactions between all particles and do not cutoff the potential at large $r$.  We evaluate the interaction between two particles as the single interaction with the nearest periodic image.  We do not include an Ewald sum over further periodic images because our box is so large that interactions with periodic images other than the nearest one are very small.

\begin{table}
\caption{Simulation Parameters, see text.  The energy per ion at simulation time $t$ is $E(t)$.}
\begin{tabular}{ll}
Parameter & Value \\
$n$ & $7.18\times 10^{-5}$ fm$^{-3}$ \\ 
$\lambda$ & 26.17 fm \\
$E(t=6\times 10^6 {\rm fm/c})$ & 328.747886 MeV \\
$E(t=151 \times 10^6 {\rm fm/c})$ & 328.747877 MeV \\
$\langle V\rangle/N$ & 328.23243 MeV \\
$T$ & 0.3436 MeV \\
\end{tabular} 
\label{tabletwo}
\end{table}

We start by evolving the system at fixed temperature by periodically re-scaling the velocities.  We adjust the temperature so that approximately half of the system remains solid and half liquid.  After a simulation time of $5\times 10^6$ fm/c we switch to evolution at constant energy and no longer re-scale the velocities.  Thus most of our simulation is in the microcanonical ensemble at fixed energy and volume.  We evolve the system at constant energy until the total simulation time is $151 \times 10^6$ fm/c.   Energy conservation is excellent.  The total energy per ion only changed by 3 parts in $10^8$ from $t=6 \times 10^6$ fm/c to $t=151\times 10^6$ fm/c, see Table \ref{tabletwo}.  The simulation was performed on an accelerated MDGRAPE-2 board \cite{mdgrape} and took approximately nine weeks. 

\section{Results}
\label{results}

In this section we first test our molecular dynamics procedure by simulating a pure system.  Then we present results for our mixture.  A 3456 ion pure system, where each ion has the same charge ($Z=29.4$) and mass, is simulated.  One half of the initial configuration is solid and the other half is liquid.  The system is evolved at constant energy for approximately 300000 fm/c.  During this time, the temperature is expected to evolve to the melting temperature because of the release of latent heat, as new solid melts or forms.  Near the end of the simulation, we evaluate the temperature as 2/3 of the kinetic energy per ion and from this we determine $\Gamma$.  We find $\Gamma=176.1\pm 0.7$.  The $\pm 0.7$ error is statistical only and does not include possible errors from finite size or non-equilibrium effects.  Our result is in good agreement with the known $\Gamma=175$ melting point of the OCP \cite{potekhin}.  This shows that our molecular dynamics procedure can accurately describe crystallization, at least for a pure system.     

Next, we calculate the latent heat by determining the potential energy difference of 3456 ion pure liquid and pure solid configurations.  The potential energy difference is equal to the latent heat if one assumes the difference in density between the phases is small.   We find the potential energy difference per ion is $0.758 \pm 0.002 T_M$, where $T_M$ is the melting temperature.   Again, the $0.002$ error is statistical only and does not include finite size effects.  Our result is in reasonable agreement with the potential energy difference for the OCP of $0.7789 T_M$ \cite{ocp}.   Our slightly lower melting temperature and latent heat may reflect screening length effects in a Yukawa fluid compared to the OCP \cite{hamaguchi}.    Alternatively, our slightly lower latent heat may reflect finite size effects for a 3456 ion system.  This latent heat is probably not an important heat source compared to the larger energy released from nuclear reactions \cite{gupta}.   

We go on to present results for our mixture with 27648 ions.  The potential energy per ion $\langle V\rangle/N$ slowly decreases with simulation time until $t\approx 70 \times 10^6$ fm/c.  This decrease may be associated with the change in composition of the two phases, see below.  Next small fluctuations are observed in $\langle V\rangle/N$ for later times that appear to be associated with fluctuations in the amount of solid phase present in the simulation.  The potential energy averaged over the last $20\times 10^6$ fm/c is given in Table \ref{tabletwo}.    The temperature is evaluated as 2/3 of the kinetic energy per ion and we find $T=0.3436$ MeV.  

The parameter $\Gamma$, Eq. \ref{gamma}, can be evaluated for a mixture of ions.  For a single ion of charge $Z_i$, the ion sphere radius $a_i$ is the radius of a sphere that contains $Z_i$ electrons,  
\begin{equation}
a_i=\Bigl[\frac{3Z_i}{4\pi \rho_{ch}}\Bigr]^{1/3}\, ,
\end{equation}
with $\rho_{ch}$ the electron density (or ion charge density).
Therefore $\Gamma_i$ for this ion is, $\Gamma_i =Z_i^2 e^2/(a_i T)$ and averaging this over a distribution of ions yields $\Gamma$ for the mixture,
\begin{equation}
\Gamma = \frac{\langle Z^{5/3}\rangle e^2}{T}\Bigl[\frac{4\pi \rho_{ch}}{3}\Bigr]^{1/3}\, .
\label{gammamix}
\end{equation}
Note that for a pure system, this equation reduces to Eq. \ref{gamma}.
Table \ref{tablethree} gives values for $\langle Z^{5/3}\rangle$ and $\Gamma$.
The value we find for our mixture $\Gamma=247$ is higher than that for a pure OCP ($\Gamma=175$).  This suggests that all of the impurities in our crystal phase have somewhat lowered its melting temperature.  However, see the discussion below about chemical separation.

\begin{figure}[ht]
\begin{center}
\includegraphics[width=3in,angle=0,clip=true] {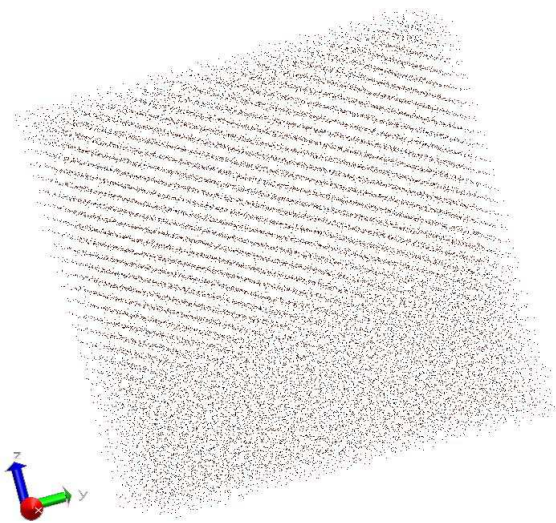}
\caption{(Color on line) Configuration of the 27648 ions at the end of the simulation. The crystal planes of the solid phase are visible in the upper half of the figure. The lower half of the figure shows a liquid phase.  The simulation volume is a cube 727.5 fm on a side.} 
\label{Fig3}
\end{center}
\end{figure}
\begin{figure}[ht]
\begin{center}
\includegraphics[width=3in,angle=0,clip=true] {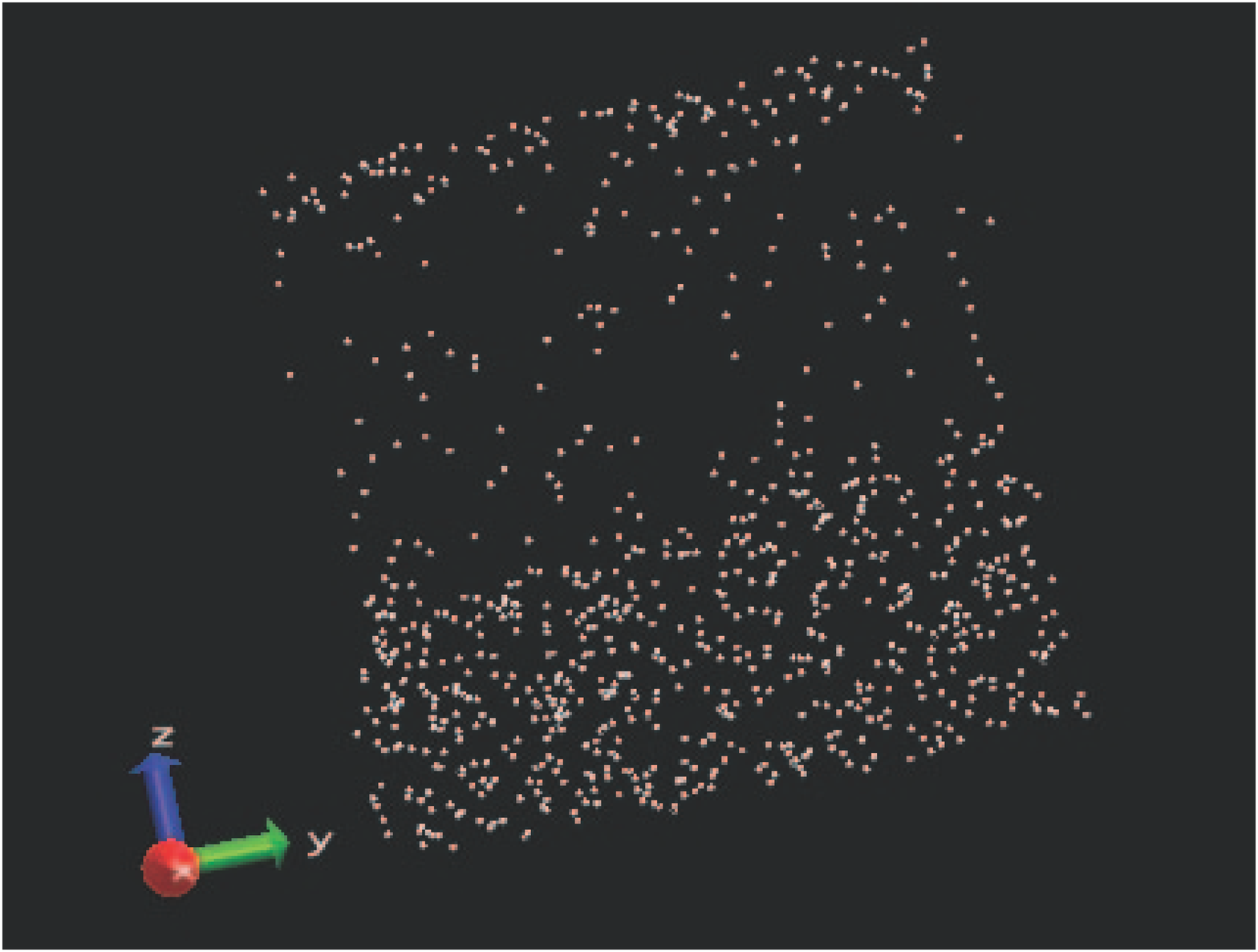}
\caption{(Color on line) Configuration of the 832 oxygen ions at the end of the simulation.  Oxygen is depleted in the solid phase, compare with Fig. \ref{Fig3}.} 
\label{Fig4}
\end{center}
\end{figure}

The configuration of the 27648 ions at the end of the simulation is shown in Fig. \ref{Fig3}.  The solid phase is visible in the upper half of the simulation volume where the crystal planes are clearly evident.  The first interface between solid and liquid is just below the center of the box and the second interface is near the top of the box.   Thus the liquid phase extends from the bottom to the top of the box because of periodic boundary conditions.  Figure \ref{Fig4} shows the final configuration of the 832 oxygen ions $Z=8$.  The oxygen ions are clearly not distributed uniformly.  Comparing Fig. \ref{Fig4} to Fig. \ref{Fig3} we see that the oxygen is greatly depleted in the solid and enriched in the liquid phase.  This directly demonstrates phase separation and shows that the composition of the liquid is different from that of the solid.

\begin{figure}[ht]
\begin{center}
\includegraphics[width=2.75in,angle=270,clip=true] {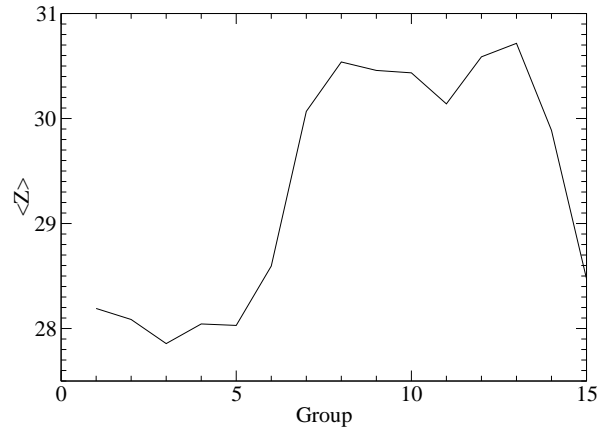}
\caption{Average ion charge $\langle Z\rangle$ in each of 15 sub-volumes.  Group 1 is at the bottom and group 15 is at the top of the simulation volume.} 
\label{Fig5}
\end{center}
\end{figure}

To further explore composition differences, we divide the ions into 15 groups according to their $z$ coordinates.  The first group includes $z$ values from 0 to $L/15$, etc.  The average charge of all of the ions in each group is plotted in Fig. \ref{Fig5}.  Groups 1-5 have a relatively small $\langle Z\rangle$ near $\langle Z\rangle\approx 28$ while groups 8-13 have a large $\langle Z\rangle\approx 30.5$.   After comparing Fig. \ref{Fig3} with Fig. \ref{Fig5}, we somewhat arbitrarily identify groups 1-5 as containing liquid phase, groups 8-13 solid phase and groups 6-7 and 14-15 as containing the two interfaces.  See Table \ref{tablethree}.  

\begin{figure}[ht]
\begin{center}
\includegraphics[width=2.75in,angle=270,clip=true] {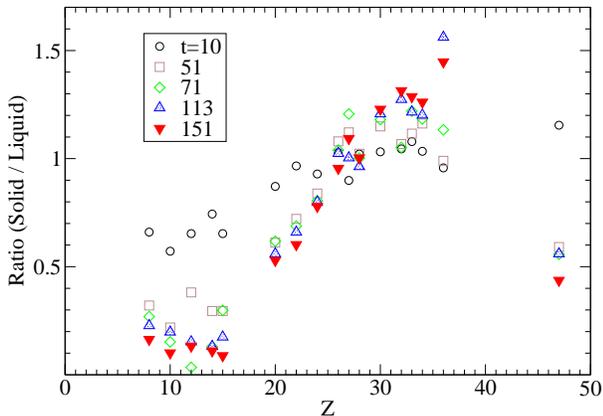}
\caption{(Color on line) Ratio of composition in the solid phase to that in the liquid phase versus atomic number $Z$, for simulation times $t$ of $10\times 10^6$ fm/c (dotted circles) to $151\times 10^6$ fm/c (solid downward pointing red triangles).} 
\label{Fig7}
\end{center}
\end{figure}

The composition of the liquid (groups 1-5) and solid (groups 8-13) are plotted in Fig. \ref{Fig1}, note the log scale, and listed in Table \ref{tableone}.  The compositions of the liquid and solid are very different.  Chemical elements with $Z\le20$ are greatly depleted in the solid phase, while most high $Z$ elements are enhanced in the solid phase.  Figure \ref{Fig7} plots the ratio of the composition in the solid to that in the liquid phase for different simulation times.  This ratio, at $t=151\times 10^6$ fm/c, is approximately linear in $Z$ for $15\le Z\le 36$.  This suggest the affinity of a given element for the solid decreases as $Z$ decreases from that of the dominant crystal species $Z=34$.  Elements with even smaller $Z<15$, while still greatly depleted in the solid, do not follow this linear trend.  Perhaps very small $Z$ ions can occupy interstitial sites in the solid in addition to replacing higher $Z$ ions at normal lattice sites.  This could enhance their concentration in the solid.  

Finally the highest charge ions $Z=47$ are, in fact, depleted in the solid.  This goes against the general rule that the solid is enriched in high $Z$ ions.  Note that there are only a few $Z=47$ ions in the simulation.  Therefore statistical errors could be large.  Perhaps this enhancement of $Z=47$ in the liquid is a non-equilibrium effect and could go away with further time evolution.  However we note that for $Z=47$ the ratio of solid concentration to that in the liquid has been decreasing with simulation time.  Therefore it may be unlikely for the ratio to change direction and finally increase with further simulation time.  Instead, the reduction in concentration of the solid may be because $Z=47$ is a much larger charge than the dominant $Z=34$ of the crystal lattice.  This large charge may fit poorly into the existing lattice and so the ions may move, instead, into the liquid phase.



We now address the important question of a further time dependence of the composition and if our simulation has reached thermodynamic equilibrium.  In Fig. \ref{Fig7} we plot the ratio of the composition of the solid to that in the liquid for different simulation times $t$.  This ratio starts at one and decreases, at small $Z$, with increasing time.  Comparing the ratio at $t=113\times 10^6$ fm/c with that at $151\times 10^6$ fm/c reveals a small but perhaps systematic difference.  However, we caution that this figure is based on our somewhat arbitrary choices of liquid and solid regions at different times.  If this difference with time is real it may suggest that the composition will continue to evolve very slowly for even larger simulation times.  This is an important open question.  In the future we will present results for longer simulation times and for simulations that start with very different compositions for the liquid and solid.  Nevertheless, we believe the ratio in Fig. \ref{Fig7} clearly shows that the liquid and solid are expected to have very different compositions. 

Finally, we discuss the charge and mass densities of the two phases, see Table \ref{tablethree}.  Within small statistical errors, we find that the charge density of the liquid is equal to that of the solid.  This implies that the number density of ions is larger in the liquid phase because the average ion charge $\langle Z\rangle$ is larger in the solid phase.  This equality of charge densities is expected in order to cancel the electron charge density.  We find that the average mass number $\langle A\rangle$ is lower in the liquid than in the solid phases.  Finally the baryon density of the liquid is slightly smaller than that of the solid phase.  Note that this small difference in density may have a significant statistical error and may be sensitive to the original distribution of $Z$ and $A$ that we use \cite{gupta}.  Neutron stars have very large gravitational fields.  Therefore, chemical separation followed by the sinking of the denser phase, can provide a significant source of heating.  Although we find only a small density difference, this should be checked in future work involving different initial compositions.  Table \ref{tablethree} also lists values of $\langle Z^{5/3}\rangle$ and $\Gamma$, see Eq. \ref{gammamix}, for the different phases.  Because $\langle Z^{5/3} \rangle$ is smaller in the liquid phase we find that $\Gamma$ is about 10\% smaller in the liquid phase compared to that in the solid phase.    

\begin{table}
\caption{Properties of the original mixture and of the final liquid and solid phases after a simulation time of $151\times 10^6$ fm/c, see text.  The impurity parameter $Q$ gives the mean square dispersion in charge, see Eq. \ref{eqq}, $\rho_{ch}$ is the ion charge density and $\rho_b$ is the baryon density. }  
\begin{tabular}{llll}
Parameter & Mixture & Liquid & Solid \\
$\langle Z\rangle$ & 29.30 & 28.04 & 30.48 \\
$Q=(\Delta Z)^2$ & 38.9 & 52.7 & 22.3 \\
$\langle Z^{5/3}\rangle$ & 285.8 & 269.0 & 301.5 \\
$\langle A\rangle$ & 87.62 & 83.8 & 91.2 \\
$\rho_{ch}$ (fm$^{-3}$) & $2.104 \times 10^{-3}$  & $2.100 \times 10^{-3}$ & $2.103 \times 10^{-3}$ \\
$\rho_b$ (fm$^{-3}$) & $6.291\times 10^{-3}$ & $6.277\times 10^{-3}$ & $6.294\times 10^{-3}$ \\
$\Gamma$ & 247 & 233 & 261 \\
\end{tabular} 
\label{tablethree}
\end{table}

\section{Discussion and Conclusions}
\label{conclusions}
How will chemical separation change the structure of a neutron star?  Consider a steady state situation where matter accretes onto a thin ocean while ocean material crystallizes to form new neutron star crust.  We assume the mass of the crust is much larger than that of the ocean.  In steady state, the rate of crystallization is equal to the accretion rate.  Furthermore, let us assume the composition of the crust is uniform.  Steady state equilibrium than requires the composition of the crust to be equal to that of the accreting material.

However, the composition of the thin ocean must become significantly enriched in light elements so that this liquid can be in thermodynamic equilibrium with the solid crust.  Note that the ocean became enriched in light elements because the first material to crystallize was depleted in light elements.  Furthermore, this initial change in composition of the crystallized material will not noticeably change the net composition of the crust because the crust is assumed to be much more massive than the ocean.    

We find a significant enrichment of oxygen $Z=8$ in the liquid.  Our original composition did not include any carbon $Z=6$ because Gupta et al. \cite{gupta} found the carbon was burned to oxygen.   However if this incorrect and carbon is present, it should also be enriched in the liquid because it has a similar atomic number to oxygen.  Therefore carbon could be significantly enriched in the ocean compared to its concentration in either the accreting material or in the crust.  Alternatively, carbon may burn, either stably or unstably, before it reaches this phase transition region.  In this case, because there is no carbon remaining, it will not be enriched in the liquid. 

Very energetic type I X-ray bursts known as superbursts \cite{superobserve,sbo2} are thought to involve unstable carbon burning.  Cumming and Bildsten argue that the mass fraction of carbon must be large, $X_{12}\approx 0.05-0.10$, in order for carbon to burn explosively \cite{superbursts,superbursts3}.  Chemical separation, which we find upon crystallization, could possibly change carbon concentrations.  In addition, chemical separation could change the thermal conductivity of the crust. This will be discussed in later work, and could impact how the ignition temperature is reached for superbursts.  In addition, the release of latent heat and or gravitational potential energy could change the temperature profile of the star.  However, the small latent heat of a pure system, that we found at the beginning of  section \ref{results}, and the small density difference between our liquid and solid phases suggests that both of these heat sources may be small.    

We find a lower melting temperature for our mixture compared to that for a one component plasma.  Table \ref{tablethree} lists $\Gamma=233$ for our liquid phase, compared to a pure one component plasma, that melts near $\Gamma=175$.  Presumably this is due to the large range of charges $Z$ that are present in our liquid phase.  This change in melting point could significantly increase the thickness of the liquid ocean in accreting neutron stars.  If the melting point does occur at $\Gamma = 233$, this implies that for accreting neutron stars with typical crust temperature $\approx 5\times 10^8\;\mathrm{K}$ that the density at which crystallization occurs is, 
\begin{equation}
\rho = 2.1\times 10^{10}\;\mathrm{g\;cm^{-3}} (T/5\times 10^8\;\mathrm{K})^3 (\Gamma/233)^3\, ,
\end{equation}
for the $\langle Z\rangle$, $\langle Z^{5/3} \rangle$ and $\langle A\rangle$ values in table III.  A rather high density of $2.1\times 10^{10}$ g/cm$^3$ for crystallization may be an order of magnitude higher than the density where $^{12}$C fuses.  Note that the frequency drifts of oscillations observed during X-ray bursts may be a way to test the depth of the crust/ ocean interface \cite{piro}.
 
However, we caution that the melting point could change if our simulation is not fully in thermodynamic equilibrium.  The phase diagram for multi-component systems can be very complicated.  For example, an additional new solid phase could form with a lower $Z$ composition after most of the high $Z$ ions have solidified.  Therefore it is important to study further the melting point of these complex mixtures. In future work we will also study phase separation for superburst ashes.

Our initial composition was not determined in a way that is consistent with chemical separation.  Gupta et al. \cite{gupta} calculated how electron capture and light particle
 reactions change the composition of rp process ash as it is compressed to higher densities.  However, they assumed the composition does not change upon crystallization.  We now find the composition changes significantly.  Therefore, one should recalculate electron capture and light particle reactions consistently with chemical separation.  We have calculated results for only one initial composition.  We expect our general result, that the liquid is greatly enriched in low $Z$ elements, to hold for a variety of different compositions.  Nevertheless, it is important to study chemical separation for other compositions.   

Itoh and Kohyama find the thermal conductivity of an impure crystal to be proportional to $1/Q$ where the impurity parameter $Q$ is the square of the dispersion in the ion charges\cite{thermalcond},
\begin{equation}
Q=(\Delta Z)^2 = \langle Z^2\rangle - \langle Z\rangle^2\, .
\label{eqq}
\end{equation}  
We find that chemical separation reduces $Q$ from 38.9 in the original mixture to 22.3 in the solid phase, see Table \ref{tablethree}.  This is because the solid contains far fewer low $Z$ ions.  {\it Therefore, chemical separation may significantly change the thermal conductivity of the crust.}  Note that $Q$ for the liquid phase is also greatly changed.  In future work we will present molecular dynamics simulation results for the static structure factor of both the liquid and solid phases and calculations of the thermal conductivity. 

The assumptions, that the composition of the crust is uniform and that the system is in steady state equilibrium, are likely to be oversimplified.  Instead the crystallization rate and composition may be time dependent.  Chemical separation could lead to the formation of layers in the crust.  There may be bands of high $Z$ material above or below bands of low $Z$ material.  This will increase the complexity of the crust and it will likely impact many crust properties.  For example, these layers could decrease the net thermal conductivity and change the temperature profile.  Alternatively, if the layers are position dependent and dynamically stable, they could change the mass quadruple moment of the star and enhance the radiation of continuous gravitational waves.  The possibility of layers should be studied in future work.

In conclusion, nucleosynthesis on the surface of accreting neutron stars likely produces a range of chemical elements.  We have performed molecular dynamics simulations of crystallization to see how this complex material forms new neutron star crust.  We find chemical separation, with the liquid ocean phase greatly enriched in low atomic number elements compared to the solid crust.  This change in composition can change many crust properties such as the thermal conductivity or shear modulus.  

\section{Acknowledgments}

We thank Dobrin Bossev, Jeremy Heyl, Gerardo Ortiz, Joerg Rottler, and Andrew Steiner for helpful discussions and acknowledge the hospitality of the Pacific Institute of Theoretical Physics where this work was started.  This work was supported in part by DOE grant DE-FG02-87ER40365 and by Shared University Research grants from IBM, Inc. to Indiana University.  Support for this work was also provided by the National Aeronautics and Space Administration through Chandra Award Number TM7-8003X issued by the Chandra X-ray Observatory Center, which is operated by the Smithsonian Astrophysical Observatory for and on behalf of the National Aeronautics Space Administration under contract NAS8-03060.

\vfill\eject

\end{document}